\documentclass[prd,showpacs,superscriptaddress,twocolumn,
floatfix,preprintnumbers,altaffilletter]{revtex4}

\usepackage{longtable}
\usepackage{graphicx}
\usepackage{amsmath,amssymb}
\usepackage{color}
\usepackage{tikz}
\usepackage{units}
\usetikzlibrary{shapes,arrows}

\newcommand{\beq}{\begin{equation}}
\newcommand{\eeq}{\end{equation}}
\newcommand{\bdm}{\begin{displaymath}}
\newcommand{\edm}{\end{displaymath}}

\definecolor{Gray}{gray}{0.9}

\graphicspath{{./plots/}}
\begin{document}

\title{Subtraction of correlated noise in global networks of gravitational-wave interferometers}

\author{Michael~W.~Coughlin}
\affiliation{Department of Physics, Harvard University, Cambridge, MA 02138, USA}
\author{Nelson~L.~Christensen}
\affiliation{Physics and Astronomy, Carleton College, Northfield, MN 55057, USA}
\author{Rosario~De~Rosa}
\affiliation{INFN, Sezione di Napoli, Complesso Universitario di Monte S.Angelo, I-80126 Napoli, Italy}
\affiliation{Universit\`a di Napoli 'Federico II', Complesso Universitario di Monte S.Angelo, I-80126 Napoli, Italy}
\author{Irene Fiori}
\affiliation{European Gravitational Observatory (EGO), I-56021 Cascina, Pisa, Italy}
\author{Mark Go{\l}kowski}
\affiliation{Department of Electrical Engineering, University of Colorado Denver, Denver, CO 80204, USA}
\author{Melissa Guidry} 
\affiliation{Department of Physics, College of William and Mary, Williamsburg, VA 23185, USA}
\author{Jan Harms}
\affiliation{INFN, Sezione di Firenze, Sesto Fiorentino, 50019, Italy}
\author{Jerzy Kubisz}
\affiliation{Astronomical Observatory, Jagiellonian University, Krakow, Poland}
\author{Andrzej Kulak}
\affiliation{AGH University of Science and Technology, Department of Electronics, Krakow, Poland}
\author{Janusz Mlynarczyk}
\affiliation{AGH University of Science and Technology, Department of Electronics, Krakow, Poland}
\author{Federico Paoletti}
\affiliation{INFN, Sezione di Pisa, I-56127 Pisa, Italy}
\affiliation{European Gravitational Observatory (EGO), I-56021 Cascina, Pisa, Italy}
\author{Eric~Thrane}
\affiliation{School of Physics and Astronomy, Monash University, Clayton, Victoria 3800, Australia}


\begin{abstract}
The recent discovery of merging black holes suggests that a stochastic gravitational-wave background is within reach of the advanced detector network operating at design sensitivity. However, correlated magnetic noise from Schumann resonances threatens to contaminate observation of a stochastic background. In this paper, we report on the first effort to eliminate intercontinental correlated noise from Schumann resonances using Wiener filtering. Using magnetometers as proxies for gravitational-wave detectors, we demonstrate as much as a factor of two reduction in the coherence between magnetometers on different continents. While much work remains to be done, our results constitute a proof-of-principle and motivate follow-up studies with a dedicated array of magnetometers.
\end{abstract}

\maketitle

{\em Introduction.}
A stochastic gravitational-wave background (SGWB) is a potential signal source for ground-based, second-generation interferometric gravitational-wave detectors such as Advanced LIGO \cite{aligo} and Advanced Virgo \cite{avirgo}. An astrophysical SGWB could be produced by objects such as compact binary coalescences, pulsars, magnetars, or core-collapse supernovae. A cosmological background could be generated by various physical processes in the early universe \cite{AbEA2009,AbEA2012s}.
Previous analyses have achieved interesting constraints on these processes \cite{AbEA2009,AbEA2012s,AaEA2014}.
In particular, with the recent discovery of a binary black-hole merger \cite{AbEA2016a}, there is a chance of observing a SGWB from these systems \cite{AbEA2016b}.

Typical searches for a SGWB cross-correlate data from two spatially-separated interferometers, where the detector noise is assumed to be Gaussian, stationary, and uncorrelated between the two interferometers and much larger than the signal. 
In the case where the noise is uncorrelated, the sensitivity of the search for the SGWB increases with time, $t_{\textrm{obs}}$, and with signal-to-noise ratio (SNR) proportional to $t_{\textrm{obs}}^{1/2}$.
Even though the interferometers are spatially separated, with Advanced LIGO consisting of detectors in Livingston, Louisiana and Hanford, Washington, and Advanced Virgo in Cascina, Italy, correlated noise between the detectors has been identified \cite{TCS2013,TCS2014}. 
Stationary noise lines, such as those from the 60\,Hz power line and 1\,Hz timing GPS noise, present at both LIGO sites, were notched in previous data analyzed \cite{LSC2007a,LSC2007b,AbEA2012s,AaEA2014}. 
Due to the increased sensitivity of second generation detectors, additional magnetic environmental correlations have also been identified \cite{TCS2013,TCS2014}. 
These correlations would contaminate the gravitational-wave data streams and thus inhibit the detection of the SGWB.
Correlated noise produces a systematic error that cannot be reduced by integration over time and therefore is a fundamental limit for SGWB searches.

Global electromagnetic fields such as the Schumann resonances are an example of environmental correlations between interferometers.
By inducing forces on magnets or magnetically susceptible materials in the test-mass suspension system, these fields are predicted to induce correlated noise in the spatially separated gravitational-wave detectors.
 
Schumann resonances are due to the very small attenuation of extremely low frequency (ELF) electromagnetic waves in the Earth-ionosphere waveguide, which is formed by the highly conducting Earth and the lower ionosphere. The ELF waves are reflected by the lower ionospheric layers at altitudes smaller than the half wavelength, enabling propagation of transverse electromagnetic waves similar to a low-loss transmission line. The attenuation increases with frequency reaching the maximum at the waveguide cut-off frequency of about 1500 Hz. In the lower part of the ELF range, the attenuation rate is particularly small: at 10 Hz it is roughly 0.25 dB/1000 km \cite{KM2013} . This enables observation of strong ELF electromagnetic field pulses (ELF transients) propagating around the world several times. 

The main source of ELF waves in the Earth-ionosphere waveguide are negative cloud-to-ground (-CG) atmospheric discharges, in which the vertical component of the dipole moment dominates and effectively generates the electromagnetic waves. An individual -CG discharge is an impulse of current associated with the charge transfer of about 2.5 C in the plasma channel that has a length of 2 to 3 km and lasts for about 75$\mu s$. A typical dipole moment (charge moment) of a discharge is about 6 C km \cite{Ra2003}. The spectrum of an impulse generated by -CG is practically flat up to the cut-off frequency of the waveguide. 
On Earth, mainly in the tropics, many thunderstorm cells are always active and produce about 50 -CG discharges per second. Since the vertical atmospheric discharges radiate electromagnetic waves in all directions, it leads to interference of waves propagating around the world. As a consequence, the spectrum of atmospheric noise exhibit resonances. The solutions to the resonance field in a lossless spherical cavity were obtained for the first time by W.O. Schumann \cite{Sch1951}. The predicted eigenfrequencies (10.6, 18.4, 26, 33.5 Hz) turned out to be much higher than the observed frequencies, which are close to 8, 14, 20, 26 Hz \cite{BW1960}, because of the dispersive character of the attenuation introduced by the ionosphere, which detunes the Earth-ionosphere cavity \cite{Sen1996}. The peaks of the Schumann resonances are relatively wide. Their quality factors for the first three Schumann resonance modes are about 4, 5, and 6, respectively. Due to the attenuation the coherence time of the field in the Earth-ionosphere cavity does not exceed 1 second.

The Schumann resonance background is a global field. The amplitude distribution of the following resonance modes depends on the time of day and year \cite{Ku2006}. The spectral density of the first resonance mode is about 1.0\,pT/$\textrm{Hz}^{1/2}$ and is different for different observers because of their location relative to the world thunderstorm centers. The daily and yearly changes in the amplitude are of several tens of percent and are related to the changes of the distance from the active thunderstorm centers (source-observer effect) and the intensity of discharges. These factors have influence on the correlation factor between fields measured in different locations on Earth and at different moments in time.

Strong atmospheric discharges, which are much less frequent, such as positive cloud-to-ground (+CG) discharges and cloud-to-ionosphere discharges associated with Sprites and Gigantic Jets, have a smaller contribution to the Schumann resonance background. Their influence on the fundamental limit for SGWB searches is negligible. However, these discharges generate ELF impulses that have very high amplitudes, so a different approach is required to analyze their influence on detection of gravitational waves.
 
The Schumann resonances are detected at high SNR with a worldwide array of extremely low frequency (ELF) magnetometers, which generally have the frequency bandwidth of 3-300\,Hz, with a sensitivity of $\approx 0.015 \textrm{pT}/\textrm{Hz}^{1/2}$ at 14\,Hz \cite{GaNi2015}.
These magnetometers can potentially be used to subtract correlated magnetic noise in gravitational-wave detectors.
In previous work \cite{TCS2013,TCS2014}, Thrane et al. estimated the effect of the correlated strain the Schumann resonances would generate for second-generation gravitational-wave detectors and in particular their effect on SGWB searches. 
In addition to exploring simple Wiener filter schemes using toy models for the gravitational-wave detector strain and magnetometer signals, they show how to optimally detect a SGWB in the presence of unmitigated correlated noise. 
In this paper, we carry out a demonstration of Wiener filtering with a goal of reducing the coherence between widely separated magnetometers (serving as proxies for gravitational-wave detectors). For our study, we use data from ultra-high-sensitivity magnetometers, which have been deployed for geophysical analyses. Using these previously deployed magnetometers, allows us access to instruments with superb sensitivity, located in very magnetically quiet locations. As shown in  \cite{TCS2014}, this is a requirement for successful subtraction. Unfortunately, the available magnetometers are not situated optimally to reproduce the subtraction scheme we envision for LIGO/Virgo. Ideally, one would want at least one pair of perpendicular magnetometers for each gravitational-wave detector. This witness pair should be far enough from the detector to avoid the local magnetic noise, but close enough that it measures a similar Schumann field as would be present at the detector. For this study, the witness sensors are very far away from the proxy gravitational-wave sensors. Despite these limitations, the work that follows is an important first step to realistic subtraction of correlated magnetic noise in gravitational-wave detectors.

In the most-ideal scenario, magnetometers would be stationed near the gravitational-wave detectors, not directly on-site so as to be directly affected by the local varying magnetic fields, but not so far away as to reduce the coherence of the magnetometers. 
This is likely to be within a few hundreds of meters of the gravitational-wave detectors.
We can test aspects of this by using existing magnetometer infrastructure, although the distances between these sites are significantly larger.
Another caveat is that among the existing magnetometers, the ones near to one another are orthogonal to one another, and so the efficacy of noise subtraction is limited.
For these reasons, the work that follows functions as a first step to realistic subtraction of correlated magnetic noise in gravitational-wave detectors.

In this work, we will use ELF magnetometers from two stations of the WERA project \footnote{http://www.oa.uj.edu.pl/elf/index/projects3.htm}.
The Hylaty ELF station is located in the Bieszczady Mountains in Poland, at coordinates $49.2^\circ$\,N, $22.5^\circ$\,E, in an electromagnetic environment with a very low level of anthropogenic magnetic field activity \cite{KuKu2014}.
The Hugo Station is located in the Hugo Wildlife Area in Colorado (USA) at $38.9^\circ$\,N, $103.4^\circ$\,W.
Both stations include two ferrite core active magnetic field antennas, one oriented to observe magnetic fields along the North-South direction, the other oriented to observe magnetic fields along the East-West direction.
These instruments are sensitive to the Schumann resonances as well as transient signals from individual high peak current  lightning discharges.  Such large discharges are often associated with so called transient luminous events that occur at stratospheric and mesospheric altitudes \cite{Pas2010}.
The magnetometers are also sensitive to atmospheric discharges, even when they have a very long continuing current phase. 
They have a lower cut-off frequency of 0.03\,Hz with the overall shape of the spectrum dominated by $1/f$ noise. 

In addition to the ELF stations, we will also use magnetic antennas at and nearby to the Virgo site. 
A temporary station was created at Villa Cristina between June 22-25, 2015 and June 29 - July 3rd, 2015.
This location is 12.72\,km southwest from Virgo, and the magnetometer was placed on the ground floor of an uninhabited house and oriented North-South.
The house was not running electricity, and the nearest location served by electricity is a small service building 500\,m away.
The house is surrounded by about a 2\,km radius of woods, and there was some excess magnetic noise induced by nearby truck transits working on logging.
In addition, there are 6 sensitive magnetic antennas located inside of Virgo experimental halls, which are ``Broadband Induction Coil Magnetometers'', model MFS-06 by Metronix. 
In the following analysis, we will take a single magnetic antenna from on-site to represent Virgo's magnetic environment, located in the North End Building along the West detector arm. The Virgo detector North arm is rotated about 20 degrees clockwise from geographic North.
The Hylaty and Hugo stations are 8900\,km away from one another. Hylaty and Villa Cristina and Hugo and Villa Cristina are 1200\,km and 8700\,km away respectively. This can be compared to the 3000\,km separation between the two LIGO sites. 


{\em Formalism.}
Typical searches for a SGWB use a cross$\mbox{-}$correlation method optimized for detecting an isotropic SGWB using pairs of detectors \cite{AlRo1999}.  This method defines a cross$\mbox{-}$correlation estimator:\newline
\begin{equation}
\hat{Y}=\int_{-\infty}^{\infty}df\int_{-\infty}^{\infty}df'\delta_T(f-f'){\tilde s}^{*}_{1}(f){\tilde s}_{2}(f'){\tilde Q}(f')
\end{equation}
and its variance:
\begin{equation}
\sigma^2_Y{\approx}\frac{T}{2}\int_{0}^{\infty}dfP_1(f)P_2(f)|{{\tilde Q}(f)}|^2,
\end{equation}
where $\delta_T(f-f')$ is the finite$\mbox{-}$time approximation to the Dirac delta function, ${\tilde s}_{1}$ and ${\tilde s}_{2}$ are Fourier transforms of time$\mbox{-}$series strain data from two interferometers, $T$ is the coincident observation time, and $P_1$ and $P_2$ are one$\mbox{-}$sided strain power spectral densities from the two interferometers. The SNR can be enhanced by filtering the data with an optimal filter spectrum $\tilde Q(f)$ \cite{Chr1992,AlRo1999}. 
Any correlated noise sources will appear in the inner product of the two strain channels, ${\tilde s}^{*}_{1}(f){\tilde s}_{2}(f')$.
We can see this by writing 
\begin{equation}
\begin{split}
{\tilde s}_{1}(f) = {\tilde h}_{1}(f) + {\tilde n}_{1}(f) + k_1(f) {\tilde m}(f) \\
{\tilde s}_{2}(f) = {\tilde h}_{2}(f) + {\tilde n}_{2}(f) + k_2(f) {\tilde m}(f)
\end{split}
\label{eq:soff}
\end{equation}
where ${\tilde h}_{i}(f)$, ${\tilde n}_{i}(f)$, and $k_i(f)$ are the gravitational-wave strain, the independent instrumental noise, and the magnetic coupling transfer function respectively. ${\tilde m}(f)$ is the correlated magnetic spectrum, which is the same at both detectors.
The effect of the local varying magnetic field on the interferometers is contained within the ${\tilde n}_{i}(f)$ terms, and in principle, on-site magnetometers can be used to monitor and subtract their effect from the data.
One metric for measuring this correlation is the coherence $c(f)$
\begin{equation}
c(f) = \frac{\overline{\tilde{s_1}(f) \tilde{s_2}(f)^*}}{\overline{|\tilde{s_1}(f)|} \overline{|\tilde{s_2}(f)}|}
\label{eq:coh}
\end{equation}
where $\tilde{s_1}(f)$ and $\tilde{s_2}(f)$ are the Fourier transforms of the two channels. 

A proposed method for mitigating magnetic noise in gravitational-wave detectors is to coherently cancel the noise by monitoring the magnetic environment. There are two potential ways this could be done.
First, one could directly correlate magnetometer and gravitational-wave strain data to calculate a \emph{Wiener filter}. This has the benefit of not relying on any magnetic coupling models, but the downside is that the Wiener filter will require very long correlation times to potentially disentangle local magnetic foreground from Schumann resonances and due to the small coherence between magnetometers and strain data.

The other option is to use magnetometers as witness sensors for magnetic noise produced at each test mass, apply measured magnetic coupling functions to strain data, and subtract these channels from the strain data. This has the benefit of not relying on weakly correlated measurements as in the case of the Wiener filter, but the downside is that there is no direct feedback on errors in the coupling model and cancellation performance. This transfer function model, which will depend on the propagation direction of the electromagnetic waves, will be difficult to compute precisely in practice also because of the correlation of the Schumann resonances between test masses, and magnetic coupling can potentially also vary with time. If, for example, the Schumann resonances produce very similar noise at two test masses, very precise estimates of the coupling are required to perform accurate subtraction (although this also implies that the Schumann resonances will produce smaller strain noise).
Otherwise, measurement errors in the coupling function could dominate the error of the Schumann noise estimate in the gravitational-wave strain channel. Both options are similar in the sense that they both rely on the use of witness sensors, with transfer functions either provided by a Wiener filter in the first case or a magnetic coupling model in the second case.

In this paper, we test noise subtraction between magnetometers as a first step towards subtraction of correlated noise from interferometer strain channels. Cancellation filters applied to magnetometers can be implemented in the time-domain as vectors of real numbers representing impulse responses and convolved with the input signal to obtain the vector to subtract from the gravitational-wave channel \cite{CoHa2014}.
The idea is to predict the correlated noise seen in the target channel, which could be either a gravitational-wave strain channel, or, as in the case presented in this study, a magnetometer, using witness sensors, which are magnetometers. 
We are only able to test certain aspects of this calculation, as a gravitational-wave interferometer has more than one test-mass with non-trivial couplings between them.
The witness sensors are used to predict and subtract the noise in the target channel.
In the case where the witness sensors have infinite SNR, only the non-correlated residual remains.

To maximize the efficacy of the filter, it is necessary to compute the filter during times of minimal local varying magnetic field activity, or equivalently, during times when the witness sensors are mostly detecting global electromagnetic noise.
The calculation of the filter coefficients depends on the autopower spectra of the input channels as well as the average correlation between channels.
The noise cancellation algorithm can then be written symbolically as a convolution (symbol $*$) \cite{Vas2001,CoHa2014}:
\begin{equation}
r(f) = y (f)-\sum\limits_{m=1}^M (a_i*x_i)(f)
\label{eq:subtraction}
\end{equation}
where $f$ is the frequency, $r(f)$ is the residual (or the cleaned data channel), $a_i$ is the correlation coefficient, $y(f)$ is the target channel, and $x_i$ are the witness channels. The convolution is defined as
\begin{equation}
a_i*x_i = \sum\limits_{i=-N}^N a[n] x[i-n],
\end{equation}
where N is the number of filter coefficients.
In this analysis, $y(f)$ corresponds to $\tilde s_{i}(f)$ from equation~\ref{eq:soff}.
One can predict the residual for a single witness sensor using the equation \cite{CoHa2014}
\begin{equation}
r(f)=\frac{1}{\sqrt{1-c(f)^2}}.
\label{eq:cohsubtraction}
\end{equation}
where $c(f)$ is the coherence calculated in equation~\ref{eq:coh}.


\begin{figure}[t]
\centering
\hspace*{-0.5cm}
 \includegraphics[width=3.5in]{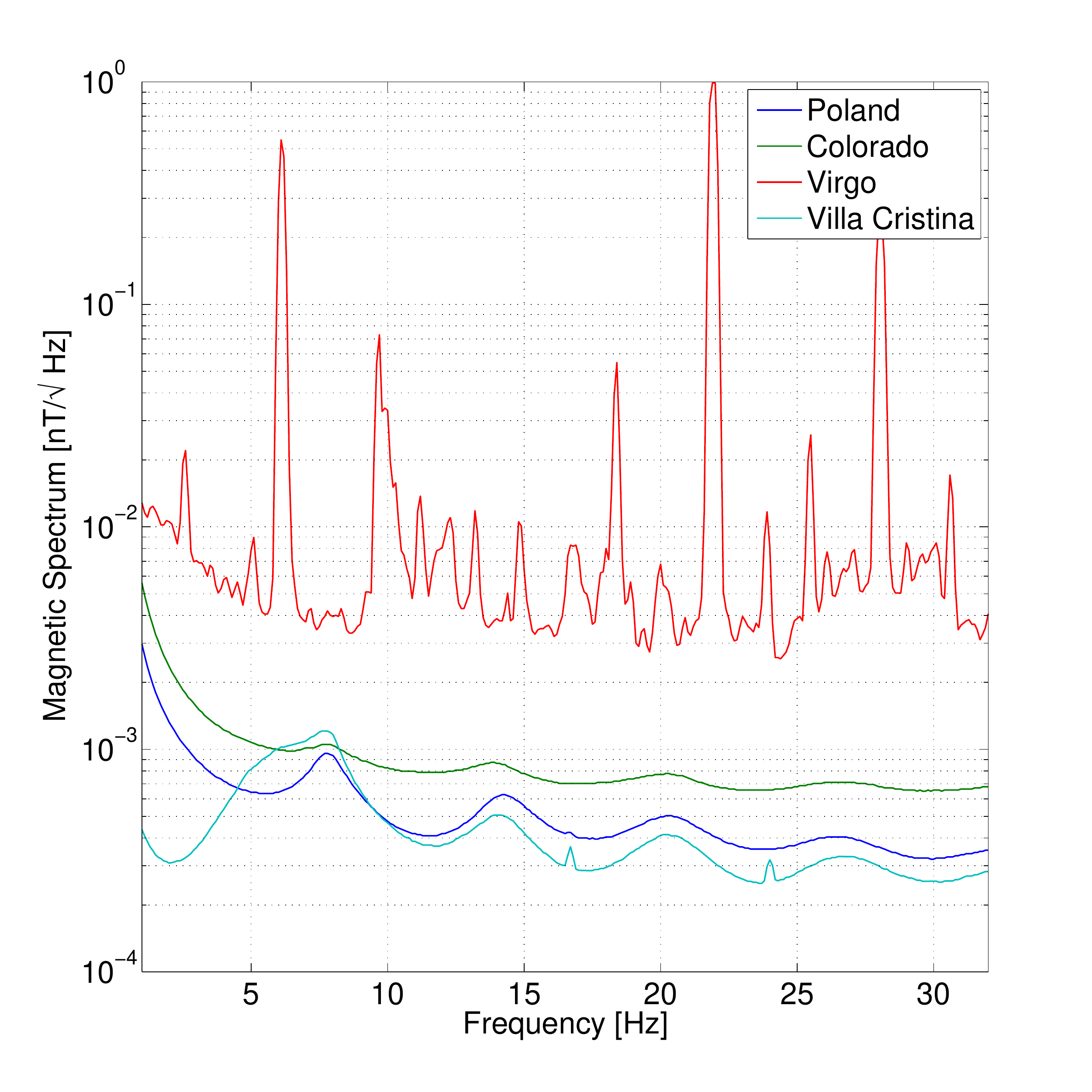}
 \caption{The median power spectral density of the North-South Poland, North-South Colorado, Virgo, and Villa Cristina magnetic antennas. These are computed using 128\,s segments. The Hylaty station in Poland, Hugo station in Colorado, and Villa Cristina antennas all achieve peak sensitivities of less than $1\,\textrm{pT}/\sqrt{\textrm{Hz}}$, while the local varying magnetic field at Virgo prevents those magnetic antennas from reaching that sensitivity.
}
 \label{fig:psd}
\end{figure}

{\em Data analysis.}
Figure~\ref{fig:psd} shows the power spectral density (PSD) of magnetic antenna data for the sites of interest in this analysis. 
The broad peaks at 8, 14, 21, 27, and 32\,Hz in the spectra correspond to the Schumann resonances, while there are many sharp instrumental line features in the Virgo spectra and two in the Villa Cristina magnetic antenna as well.
It is clear that the ELF stations contain sufficient sensitivity to detect the Schumann resonances at high SNR.
On the other hand, the magnetic antenna stationed inside of Virgo is dominated by local varying magnetic fields as well as the 50\,Hz power-line and associated sidebands.
For these local magnetic antennas, suppression of the power-line noise will be important to maximize the potential subtraction of the underlying Schumann resonances.
We can use this power-spectrum, in addition to the most recent magnetic coupling function published in \cite{AbEA2016c}, to determine the effect the global electromagnetic noise has on SGWB searches.
We note here that the calculation of common and differential mode coupling performed are conservative and in the future, separate coupling functions for each test mess would be ideal  \cite{TCS2014}.

We can relate the strain noise induced by the magnetic fields, $S_{\rm MAG}(f)$, which is the magnetic power spectrum multiplied by the magnetic coupling function, to $\Omega_{\rm MAG} $ by \cite{AbEA2012s}
\begin{equation}
\Omega_{\rm MAG} = \frac{10 \pi^2} {3 H_0^2} S_{\rm MAG}(f) f^3
\end{equation}
where we assumed a value $H_0=67.8\,\rm km/s/Mpc$ for the Hubble constant \cite{AdEA2013}.
The Schumann resonances induce correlated noise such that $\Omega_{\rm MAG} = 1 \times 10^{-9}$, which is a potential limit for Advanced LIGO.
Here we have integrated over a year at design sensitivity and included the Schumann frequency band. 
We note here that the Schumann resonances are strongest in the fundamental and first few harmonics and as well as the fact that the magnetic coupling is the strongest at the lowest frequencies \cite{AbEA2016c}.
For these reasons, if instead of performing a SGWB search from 10\,Hz, one integrates from 25\,Hz, the effect from the Schumann resonances on SGWB searches decreases by about an order of magnitude.


\begin{figure*}[t]
\centering
\hspace*{-0.5cm}
 \includegraphics[width=3.5in]{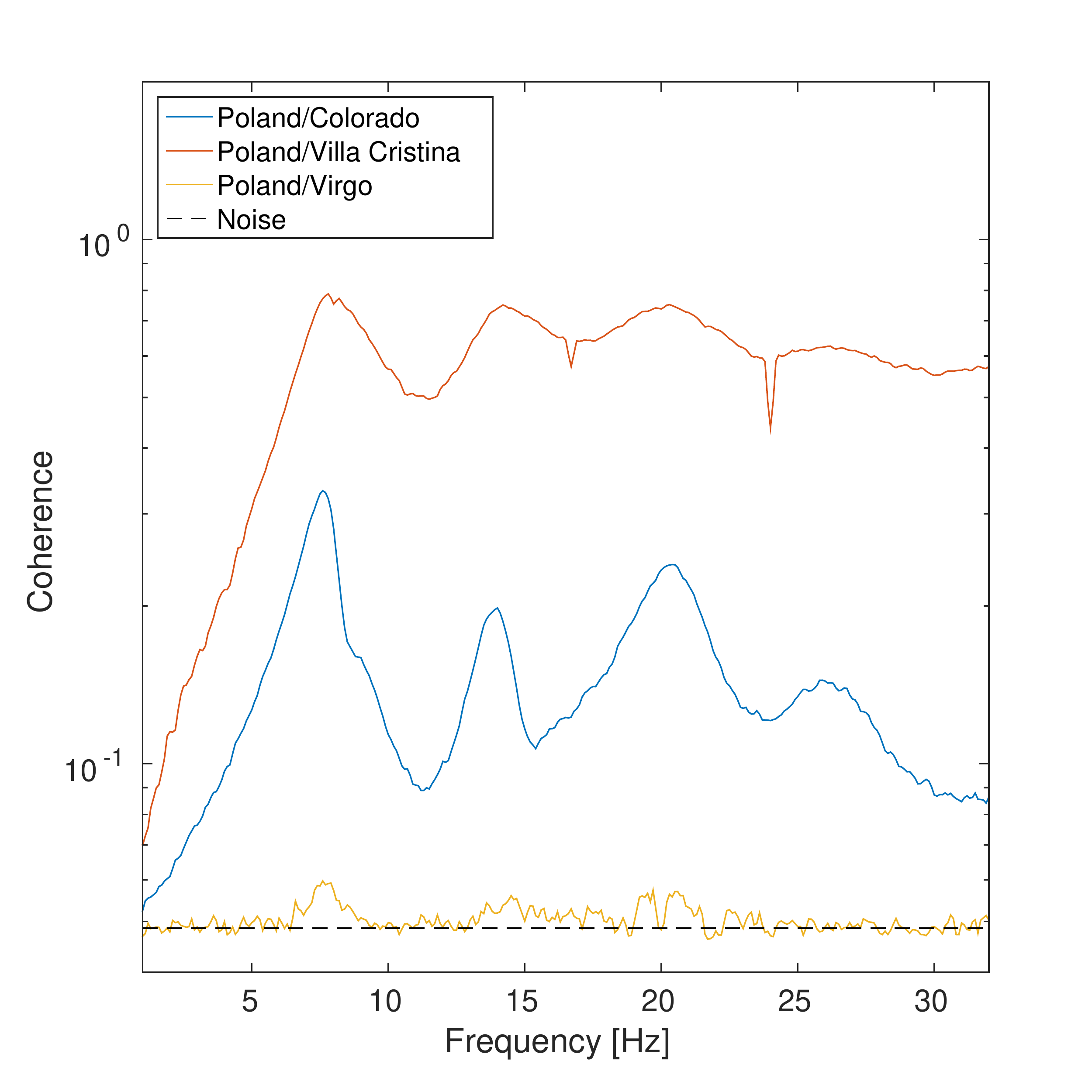}
 \includegraphics[width=3.5in]{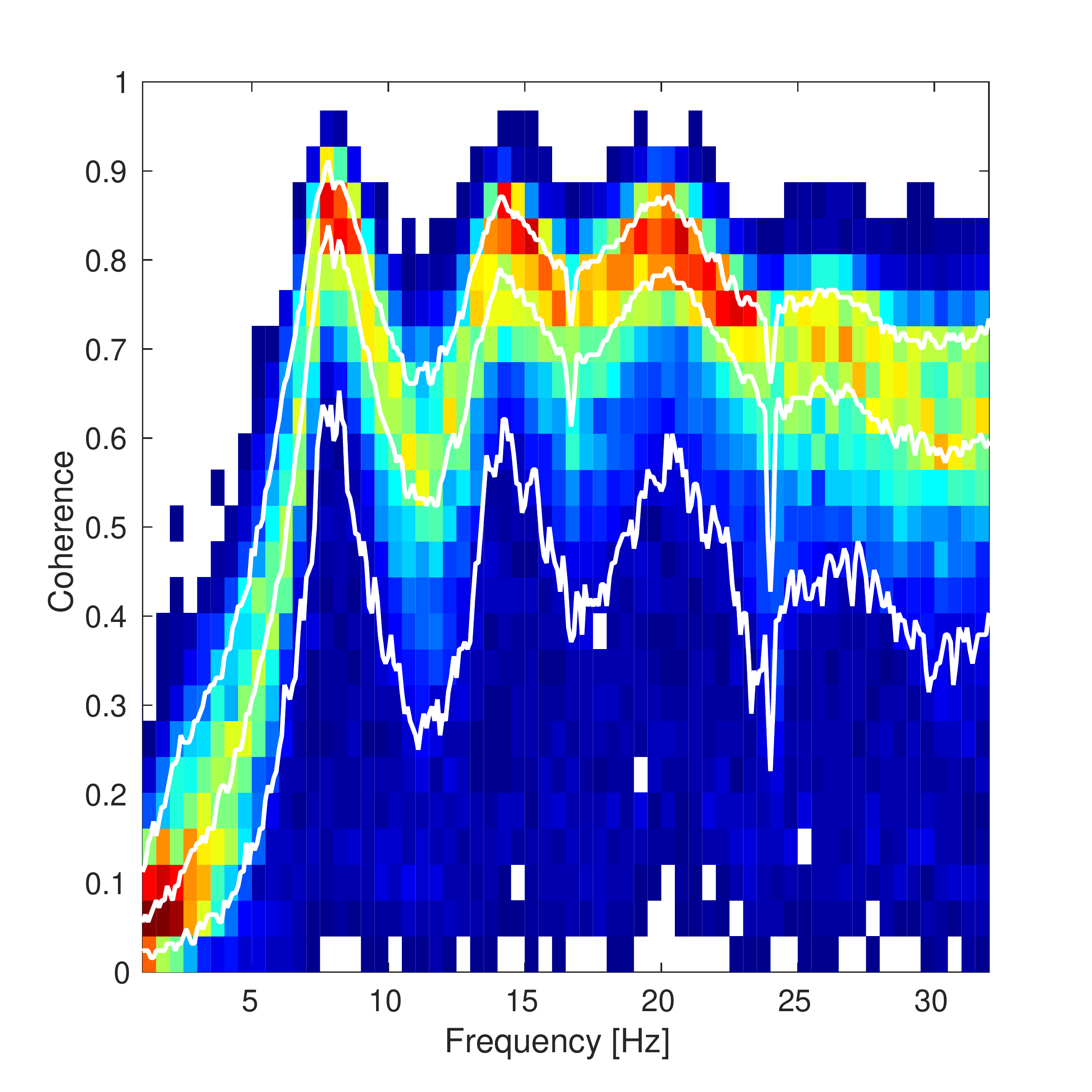}
 \caption{
The plot on the left is the coherence between the North-South Poland, North-South Colorado, Virgo and Villa Cristina magnetic antennas over 4 days of coincident data divided into 128\,s segments. In addition, we plot the expected correlation given Gaussian noise.
The coherence between the Virgo and Villa Cristina antennas is similar to that of Virgo and North-South Poland.
The plot on the right is the variation in the coherence between the Hylaty station in Poland and Villa Cristina.
The colors represent the percentage of the segments which have any given coherence value.
The white lines represent the 10th, 50th, and 90th coherence percentiles.
}
 \label{fig:coh}
\end{figure*}

The left of figure \ref{fig:coh} shows the coherence between the North-South Poland and North-South Colorado, Virgo, and Villa Cristina magnetic antennas. 
The coherence between Poland and Colorado and Villa Cristina show clear peaks in the coherence spectrum, while the coherence between the Virgo and Poland stations are less pronounced due to very high local magnetic noise level at the Virgo site.
The right of figure \ref{fig:coh} shows the variation in the coherence between the Hylaty station in Poland and Villa Cristina stations. 
These coherent peaks are potentially problematic for the SGWB searches.
The most important peaks for the gravitational-wave detectors are the secondary and tertiary harmonics at 14\,Hz and 20\,Hz respectively, as the primary is below the seismic wall of the gravitational-wave detectors at 10\,Hz.
In the future, it will be beneficial to measure the coherence between quiet magnetometers stationed near LIGO and Virgo.
As we show below, high coherence indicates the potential for significant signal subtraction.
\begin{figure*}[t]
\centering
\hspace*{-0.5cm}

 \includegraphics[width=3.5in]{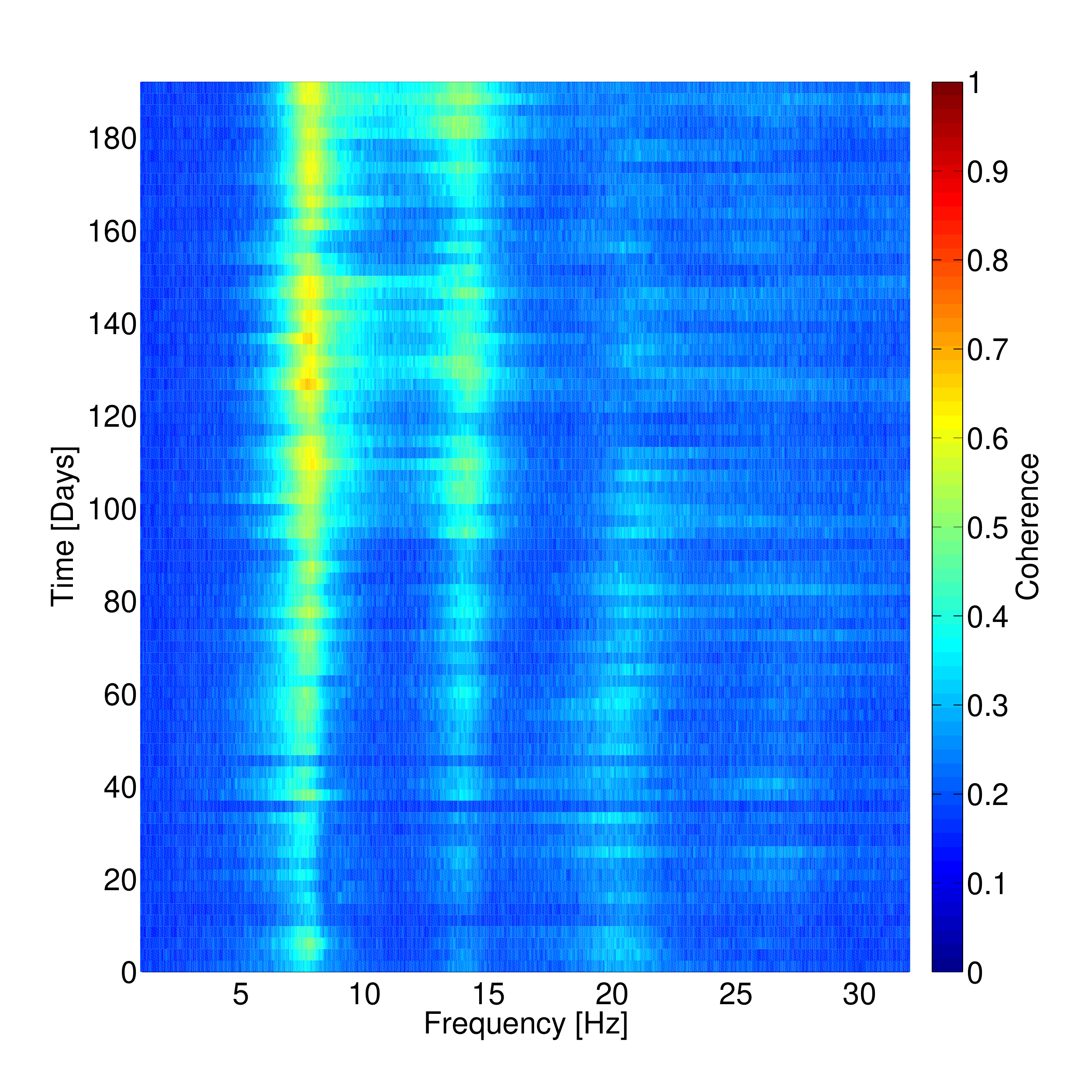}
 \includegraphics[width=3.5in]{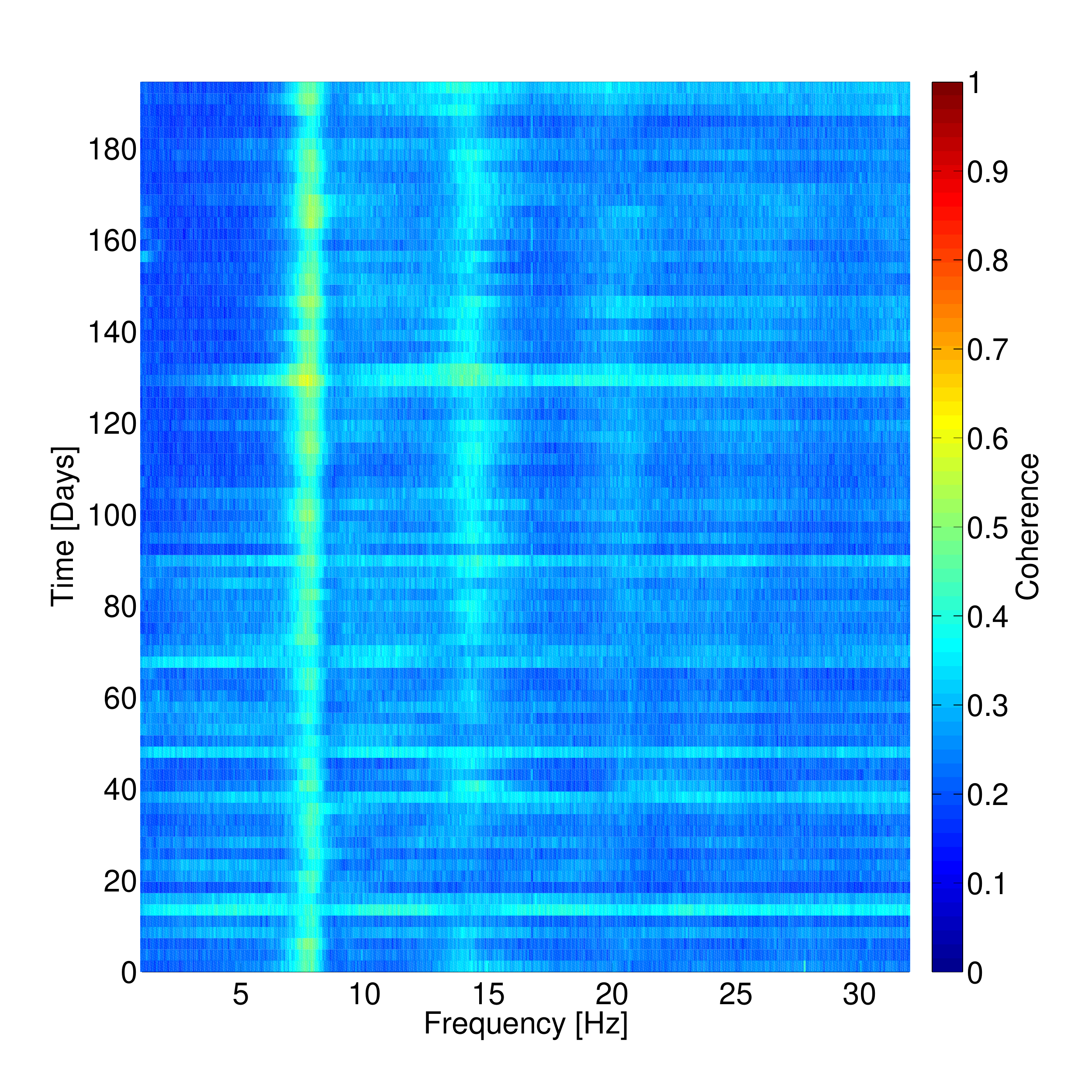}
  
 \caption{
The plot on the left is a time-frequency map of the coherence between the North-South Poland and North-South Colorado magnetic antennas. The plot on the right is the same between the North-South and East-West Hylaty station in Poland magnetic antennas. 
}
 \label{fig:cohtf}
\end{figure*}

Figure~\ref{fig:cohtf} shows the time-frequency coherence between the North-South Poland and North-South Colorado stations on the left and the coherence time-frequency map between two co-located but orthogonal Hylaty station in Poland magnetic antennas on the right.
These plots indicate that the Schumann resonances are only present, at least at detectable levels, at certain times and not others.
The coherence between the two perpendicular Hylaty station in Poland magnetic antennas show relatively weak coherence due to the fact that they overlook different regions of the thunderstorm activity on Earth.
In addition, their correlation is dominated by collocated magnetic fields, predominantly due to nearby thunderstorms, which can lead to ADC saturations.

Typically, Wiener filters are used to make a channel less noisy, e.g., \cite{CoHa2014}. The success of the filtering procedure is determined based on the observed reduction of the channel's auto power spectral density. However, when the goal is to minimize correlated noise in two channels, we must use a different metric to determine the efficacy. In the case of Schumann resonances, the noise in question contributes a very small amount to the auto power spectral density in each gravitational-wave strain channel, perhaps 0.1\%. The key metric for our study is the reduction in coherence between the two channels.

\begin{figure*}[t]
\centering
\hspace*{-0.5cm}
 \includegraphics[width=3.5in]{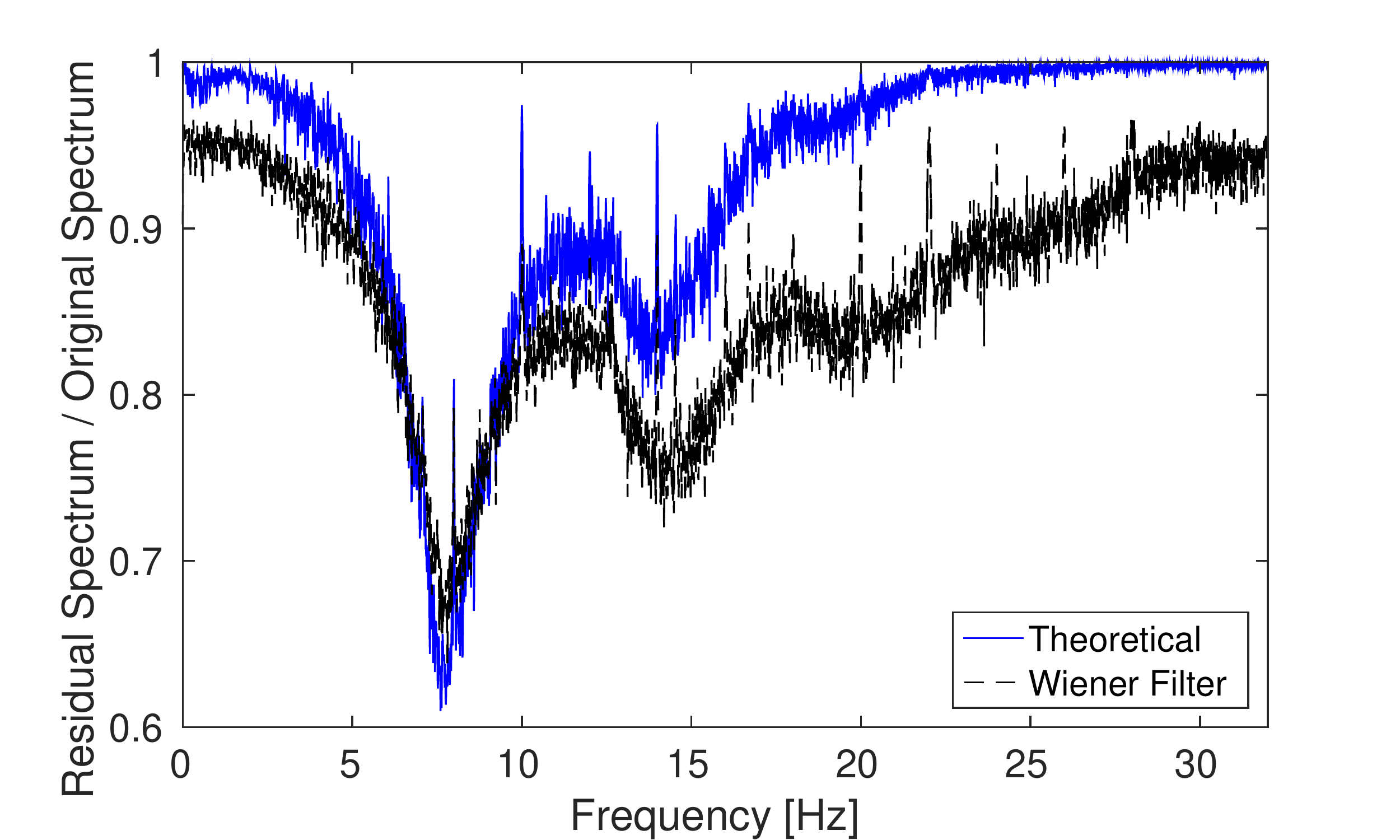}
 \includegraphics[width=3.5in]{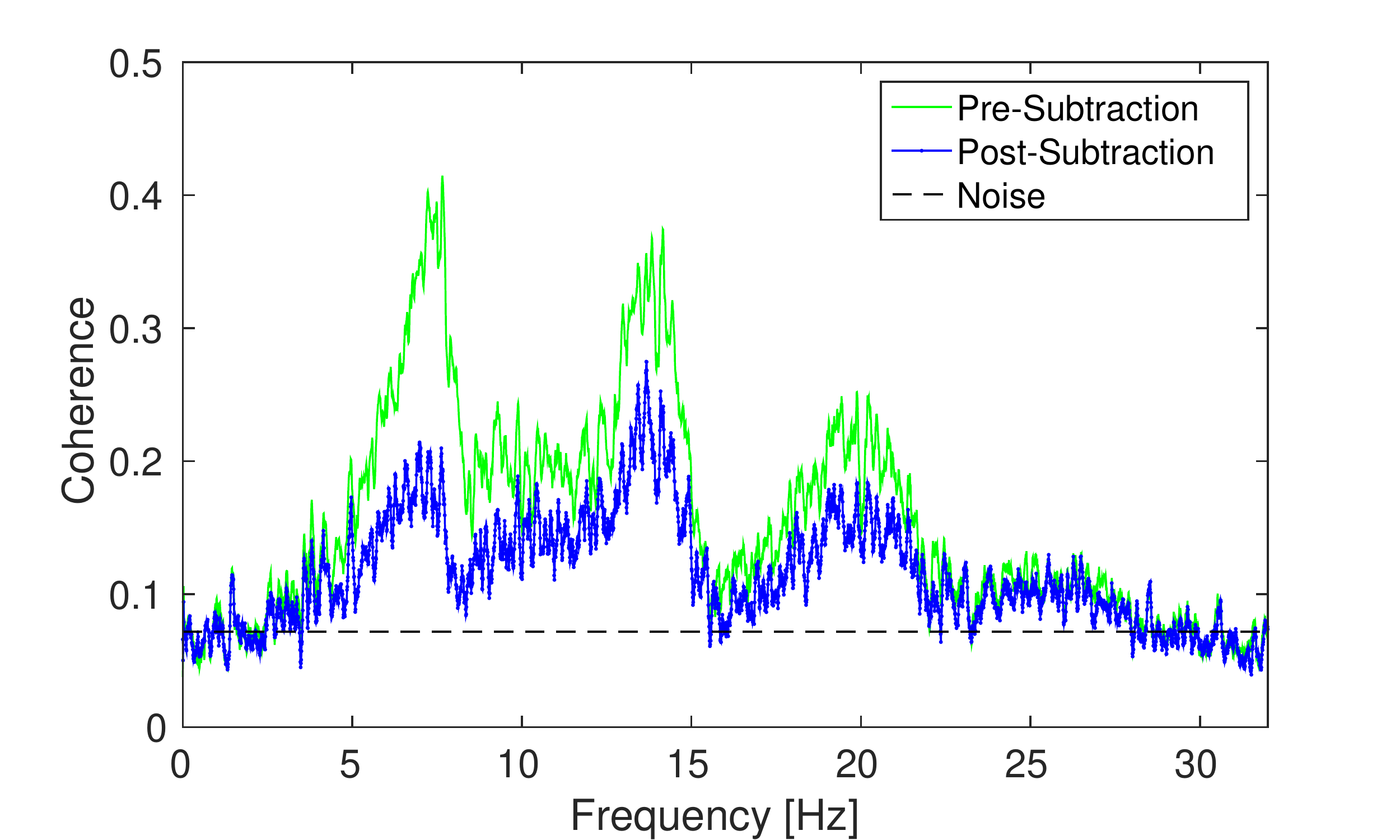}
 \caption{Success of the Wiener filtering procedure under two metrics: reduction in auto power spectral density and reduction in coherence. 
 On the left is the ratio of the auto power spectral density before and after Wiener filter subtraction using the North-South Poland antenna as the target sensor and East-West Poland magnetic antennas, the two Colorado magnetic antennas, and the Villa Cristina magnetic antenna as witness sensors. The ``theoretical'' line corresponds to the expected subtraction from the maximum coherence between the target and witness sensors (in any given frequency bin) given by equation~\ref{eq:cohsubtraction}, while the ``subtraction'' line corresponds to the subtraction achieved by the Wiener filter.
 The ``subtraction'' line exceeds the performance over the ``theoretical'' over part of the band because the theoretical line, computed from a single channel, ignores the benefit of having multiple witness sensors.
 On the right is the coherence between the North-South Colorado and the Villa Cristina magnetic antenna before and after Wiener filter subtraction using the North-South Poland antenna. 
 There is approximately a factor of 2 reduction in coherence at the fundamental Schumann resonance.}
 \label{fig:subtraction}
\end{figure*}

As described above, Wiener filtering allows for the subtraction of noise from a particular target channel using a set of witness channels. 
It is instructive to think of the target magnetometer as a proxy for a gravitational-wave interferometer strain channel. 
The above analysis is similar to the potential scenario in subtraction of the Schumann resonances from gravitational-wave interferometer strain channel data. In this case, both on-site and off-site magnetometers will be used in coordination to subtract the effects of both local and global electromagnetic fields from the strain data.
If the local varying magnetic fields have a difference in phase and/or direction with the global magnetic fields, their use will likely inject extra noise from the Schumann resonances when subtraction is performed.
In our particular example, we take the North-South Poland magnetic antenna as our target channel and the East-West Poland magnetic antenna, the two Colorado magnetic antennas, and the Villa Cristina magnetic antenna as the witness channels. 
Using both of the magnetometers allows for coverage of both the North-South and East-West magnetic fields.
The idea is that using closely spaced, orthogonal magnetometers are likely to contain complementary information, even if local varying magnetic field disturbances will be correlated between them.
On the other hand, using co-located magnetometers measuring magnetic fields in the same direction, assuming the local varying magnetic fields dominate, provides higher SNR for the Wiener filter if and only if the intrinsic noise (not due to the global electromagnetic noise) in each magnetometer is independent.
This example shows the subtraction that can maximally be expected given the currently deployed sensors in this study and properties of the local varying magnetic fields. 
It is also similar to the gravitational-wave detector case in that the magnetometers used to perform coherent subtraction will be a combination of local magnetometers to subtract the local varying magnetic fields and magnetometers installed outside of the gravitational-wave detector beam arms to maximize coherence with the Schumann resonances; this has the benefit of limiting the effects from the local varying magnetic fields on site, which can be quite strong \cite{AaEA2012}.
In this case, using two witness sensors subject to the same local varying magnetic fields is suboptimal, as any noise in the witness sensors limits the efficacy of the subtraction. 
In the case where the witness sensors have the same limiting noise source, which is generally a sum of instrumental and environmental/local noise sources, the effective coherence will be dominated by the local varying magnetic fields. 
This is true regardless of the target sensor, be it another magnetometer or a gravitational-wave detector strain channel.
This noise floor will set the possible SNR for Wiener filtered subtraction, which must be high enough to provide the required subtraction.

On the left of figure~\ref{fig:subtraction}, we show the subtraction using a frequency domain Wiener filter between 1-32\,Hz where we use 30\,minutes of data to generate the filter.
In this analysis, we use a time with minimal local disturbances to estimate the Wiener filter and apply it across the entire run. 
This makes an implicit assumption that the correlation between the target and witness sensors is not changing significantly with time.
We have checked that updating the Wiener filter every five minutes gives similar results, indicating that the time-scale does not make a significant difference in this case for the magnetic field subtraction.
Magnetometers dominated by local varying magnetic fields may require regular updates if the local varying magnetic field is changing often such that the sensitivity between the gravitational-wave detectors and magnetometers to the Schumann resonances changes.
In the case where the target sensor is a gravitational-wave strain channel, it will likely be useful to regularly update the filter due to possible changes in the magnetic coupling function of the detector with time.
The time between updates will be affected by interferometer commissioning activities interspersed with data acquisition.
The coherence results discussed above give us an expectation of the amount of subtraction we can expect between magnetometers using equation~\ref{eq:cohsubtraction}. 
This is consistent with the result of the Wiener filter implementation.

We now turn our attention to the metric most appropriate for searches for stochastic gravitational-wave backgrounds.
We can use the available magnetic antennas as a proxy for a 2-detector for a gravitational-wave interferometer network.
We assign the North-South Colorado and Villa Cristina magnetic antennas to be gravitational-wave strain channels and use the North-South Poland magnetic antenna to subtract the coherent Schumann resonances. 
On the right of figure~\ref{fig:subtraction}, we measure the coherence between the North-South Colorado and Villa Cristina magnetic antennas before and after the subtraction, to measure the effect that the Wiener filtering has had on the correlations.
We find a reduction of approximately a factor of 2 in coherence near the peak of the dominant harmonic.

Using the results of Thrane et al. \cite{TCS2014}, we can place these results in context for Advanced LIGO. The authors of that work showed that the integrated SNR from correlated magnetic noise in one year of coincident data from the LIGO Hanford and Livingston detectors operating at design sensitivity is between 24-470, depending on magnetic field coupling assumptions, significantly limiting a potential measurement of $\Omega_\textrm{GW}$. With the help of recent commissioning activities to improve the magnetic coupling functions, these numbers are likely to be a worst-case-scenario. The idea is that correlated noise can only be safely ignored in SGWB searches if the SNR contribution from correlated noise is much less than 1. As $\textrm{SNR} \propto {\tilde r}^{*}_{1}{\tilde r}_{2} {\tilde m}^{*}_{1}{\tilde m}_{2}$, any reduction made in the power spectrum of the magnetic noise ${\tilde m}^{*}_{1}{\tilde m}_{2}$ will reduce the SNR by that same factor.
One possibility to improve upon this subtraction would be to use multiple sensors to improve the effective SNR of the witness sensors. 
Another (perhaps more promising) possibility would be to use magnetometers located closer to the gravitational-wave interferometer site.


{\em Conclusion.}
In summary, the magnetic fields associated with Schumann resonances are a possible source of correlated noise between advanced gravitational-wave detectors.
The optimal method for subtracting correlated noise in these detectors is Wiener filtering.
In this paper, we have described how the global electromagnetic fields create a potential limit for SGWB searches with advanced gravitational-wave detectors. In particular, without subtraction, the Schumann resonances induce correlated noise such that $\Omega_{\rm MAG} = 1 \times 10^{-9}$, using the most recent magnetic coupling function published in \cite{AbEA2016c} and neglecting common-mode rejection, is a potential limit for Advanced LIGO, where we have integrated over a year at design sensitivity and included the Schumann frequency band. 
We have also discussed the implications of the coherence achieved between extremely low frequency magnetometers in their use in stochastic searches. 
This coherence is sensitive to the magnetometer SNR of the Schumann resonances to the fundamental instrument noise as well as the local varying magnetic fields. 
Both the LIGO and Virgo sites will benefit from sensitive magnetometers at magnetically quiet locations that are outside of the buildings housing the gravitational-wave detectors. 
We have also shown that careful treatment of the magnetometer data, which include significant sensitivity fluctuations due to local varying magnetic fields, will be required for subtraction of the Schumann resonances.
We show that magnetometer pairs thousands of kilometers apart are capable of reducing magnetic correlations by about a factor of 2 at the fundamental peak.
This gives hope that magnetometers near to the interferometers can effectively subtract magnetic noise with Wiener filtering. 

There is significant work to be done looking forward.
It will be a challenge to measure the Schumann resonances at the sites due to the local magnetic foreground.
It will be important to perform a similar measurement where the magnetometers are aligned and separated by a 3 to 4\,km distance, which will more closely mimic the scenario when magnetometers are used as witness sensors for gravitational-wave detectors.
Finally, it will be useful to have coupling measurements at all of the test masses at the gravitational-wave interferometers, which will determine the level of the correlations of the Schumann resonances at multiple test masses.


MC is supported by the National Science Foundation Graduate Research Fellowship Program, under NSF grant number DGE 1144152.
NC work is supported by NSF grant PHY-1505373.
ET is supported through ARC FT150100281.

\bibliographystyle{unsrt}
\bibliography{references}

\end{document}